# Is There Quantum Gravity in Two Dimensions?[1]

Wolfgang Beirl[2] and Bernd A. Berg[3,4]

(wb@kph.tuwien.ac.at, berg@hep.fsu.edu)

**Abstract**

A hybrid model which allows to interpolate between the (original) Regge approach and dynamical triangulation is introduced. The gained flexibility in the measure is exploited to study dynamical triangulation in a *fixed* geometry. Our numerical results support KPZ exponents. A critical assessment concerning the apparent lack of gravitational effects in two dimensions follows.

[1]This research was partially funded by the U.S. Department of Energy under contracts DE-FG05-87ER40319, DE-FC05-85ER2500, and by the Austrian "Fonds zur Förderung der wissenschaftlichen Forschung" under contract P9522-PHY.

[2]Institut für Kernphysik, Technische Universität Wien, A-1040 Vienna, Austria.
[3]Department of Physics, The Florida State University, Tallahassee, FL 32306, USA.
[4]Supercomputer Computations Research Institute, Tallahassee, FL 32306, USA.

# 1  Introduction

Two numerical methods, both derived from Regge calculus, have been used to investigate non-perturbative quantum gravity: dynamical triangulation (DT) for fixed, equal link lengths, called standard dynamical triangulation (SDT) henceforth, and the (original) Regge approach (RA), varying the link lengths of a simplicial lattice with fixed incidence matrix. The latter suffers from the same fundamental problem as continuum formulations: the correct measure is unknown. On the other hand, the SDT approach is claimed to be free of those troubles since the summation of triangulations is thought to provide the unique, correct measure of geometries. In the following we focus on 2D Ising spins coupled to the geometry, as in this case a number of exact results are known.

For 2D Ising spins the SDT method is known to be equivalent to exactly solvable matrix models [1, 2]. The fact that KPZ theory [3] reproduces precisely the the same critical exponents is widely regarded as a "proof" [4] that the SDT method is correct (not only in 2D). In contrast the RA has failed to obtain the KPZ results [5, 6]. Instead consistency with Onsager's exponents for the flat space Ising model was found. Recently, the scenario has become further diffused by the claim of Vekić et al. that KPZ exponents are observed for a *flat* space Ising spin system, where variable incidence matrices are used [7].

Let us introduce a hybrid model which allows to interpolate between RA and SDT. The path integral reads

$$Z_{N_0}(\beta, \lambda, a) = \sum_{\{T\}} W_T \int D\{l^T\} \sum_{\{s\}} e^{-E} \tag{1a}$$

with

$$E = \sum_i \left( \lambda A_i^T + a \frac{(\delta_i^T)^2}{A_i^T} \right) + \frac{\beta}{2} \sum_{l^T} A_{l^T} \left( \frac{s_{i(l^T)} - s_{j(l^T)}}{l^T} \right)^2 \tag{1b}$$

Here $s_i = \pm 1$, $i = 1, ..., N_0$ are the usual Ising spins, and $i, j$ label sites of the lattice. The first sum is over distinct triangulations $T$ of a given topology, weighted with $W_T \geq 0$. The adjacency matrix is defined by

$$T_{ij} = \begin{cases} 1, & \text{if } i \text{ and } j \text{ are neighbours,} \\ 0, & \text{otherwise.} \end{cases}$$

Now $l^T$ labels the links $l_{ij}$ for which the adjacency matrix $T_{i(l)j(l)}$ is non-zero, where $i(l)$ and $j(l)$ are the sites connected by $l$. We use $t^T$ to label the triangles of the triangulation $T$. The area $A_i^T$, associated with a site $i$ is taken to be baricentric

$$A_i^T = \frac{1}{3} \sum_{t^T \supset i} A_{t^T},$$

where $A_{t^T}$ denotes the area of the triangle $t^T$. By $n_i$ we denote the number of triangles encountered in the sum, *i.e.* the coordination number. The deficit angle at site $i$ is defined as

$$\delta_i^T = 2\pi - \sum_{t^T \supset i} \theta_i(t^T),$$

where $\theta_i(t^T)$ is the dihedral angle associated with the triangle $t^T$. Similarly, the area $A_{l^T}$, associated with the link $l^T$, is defined as

$$A_{l^T} = \frac{1}{3} \sum_{t^T \supset l^T} A_{t^T}.$$

Finally, we use $N_1$ and $N_2$ to denote the numbers of links and triangles, respectively. Choosing $W_T$ and $D\{l^T\}$ to interpolate between RA and SDT, we expect that the model can shed new light on the reasons for the occurance of different universality classes, characterized by Onsager versus KPZ exponents. For the RA all weight factors but one are zero, whereas for DT all weight factors are identical, for instance $W_T \equiv 1$. In this paper we focus on DT with the measure

$$D\{l^T\} = \prod_{ij} T_{ij} dl_{ij} \delta(l_{ij} - c_{ij}), \tag{2}$$

where $(c_{ij})$ is a $N_0 \times N_0$ matrix of constant link lengths. With

$$c_{ij} \equiv 1 \tag{3a}$$

the SDT approach is obtained. Another notable choice is

$$c_{ij} = \text{geodesic distances between vertices } i, j \text{ in a fixed geometry.} \tag{3b}$$

With (3b) it is easy to see that the geometry (of the starting configuration) stays fixed. Variations due to changes in the incidence matrix $T_{ij}$ are compensated by appropriate changes of the link lengths. Clearly, the general measure in (1a) allows to interpolate smoothly between (3a) and (3b), for instance by giving the two length assignments different weights which add to one. Such a parametrization allows to study the onset of gravitational dressing when proceeding from (3b) to (3a). Correspondingly, one may expect the critical exponents to vary between Onsager for (3b) and KPZ for (3a). However, in the next section we present simulation results with the measure (3b), already supporting KPZ exponents. The DT scenario becomes thus very similar to the RA results reported in [5, 6]: Turning "gravity" on or off seems to have no effect on the critical exponents. A critical assessment of these results is tried in section 3. Summary and conclusions are given in section 4.

## 2 Simulations

We present simulation results with the measure (3b). The action is given by (1b) with $a = 0$. Since we consider ensembles of fixed area the parameter $\lambda$ is of no significance. As in previous simulations [8, 9, 10, 11] with the 2D DT approach we rely on the spherical topology.

### 2.1 Algorithm

We first describe how both $T_{ij}$ and $c_{ij}$ can be generated through a Metropolis algorithm. We start with a geometry specified by a triangulation $T_{ij}^0$ with equilateral triangles:

$$T_{ij}^0 = 1 \Rightarrow c_{ij} = 1. \tag{4}$$

All other triangulations are generated as for SDT by applying successively link-flip moves. In difference to SDT we change the lengths of the flipped links to get $c_{ij}$ as defined by (3b). The procedure is as follows: The two triangles affected by the link flip are embedded in $R^2$ as shown in figure 1. The vertices are labelled as $i, j, k, l$. The length $c_{ij}$ of the flipped link is calculated from the known lengths, $c_{ij} = c_{ij}(c_{ik}, c_{il}, c_{jk}, c_{jl})$, using the Euclidean distance

in flat space

$$x_1 = \frac{c_{kl}^2 - c_{il}^2 + c_{ik}^2}{2c_{kl}}, \ y_1 = \sqrt{c_{ik}^2 - x_1^2}$$
$$x_2 = \frac{c_{kl}^2 - c_{jl}^2 + c_{jk}^2}{2c_{kl}}, \ y_2 = \sqrt{c_{jk}^2 - x_2^2}$$
$$c_{ij} = \sqrt{(y_1 + y_2)^2 + (x_1 - x_2)^2}.$$

It is easily seen, that all (geodesic) distances, the deficit angles and the (total) area are preserved by this flip. The geometry stays unchanged. Since link flips may successively lead to large lengths $l_{ij}$ it is necessary to use a cutoff $l_C$ so that only configurations with $l_{ij} < l_C$ are allowed. Also the algorithm has to reject certain flips to enforce that the area stays invariant, see figure 1. In addition we reject flips which lead to triangles with area zero.

In the next subsection we present numerical results for two different geometries. To construct $T_{ij}^0$, we begin with the simplest triangulation of the 2-sphere: the surface of a tetrahedron. This provides for a list of $N_0 = 4$ vertices, $N_1 = 6$ links and $N_2 = 4$ triangles together with incidence tables describing how they are connected. Then one divides successively the links applying Alexander [12] moves, which increase $N_0$ by one, $N_1$ by three and $N_2$ by two. After each move the new simplices are added at the end of the incidence tables and the procedure is iterated until the desired number of vertices is reached. Increasing the index of the link which is split sequentially gives geometry $G1$, while increasing it in steps of two results in $G2$. Cutoffs $l_C^1 = l_C^2 = 5.4$ are used for both geometries. For $N_0 = 250$ the distributions of coordination numbers $n$ are given in table 1. For larger $N_0$ the distribution of the coordination numbers converges fast towards a fixed shape, which allows to speak of fixed geometries.

## 2.2 Finite size scaling preliminaries

For both geometries we have performed Metropolis simulations to estimate critical exponents through finite size scaling (FSS) analysis. Similar as in [6] we define pseudocritical points $\beta_c^i(N_0)$, $i = 1, 2, 3, 4$ as maxima of suitable physical quantities. They are:

1. The magnetic susceptibility

$$\chi(\beta) = N_0 \left( <m^2> - <|m|>^2 \right), \qquad (5)$$

where $m$ is the lattice average $m = N_0^{-1} \sum_i s_i$.

2. The derivative of the absolute value of the magnetization:

$$D^m(\beta) = \frac{d<|m|>}{d\beta} = <E><|m|> - <E|m|>. \qquad (6)$$

3. The derivative of the Binder [13] cumulant $B = 1 - \frac{1}{3} <m^4><m^2>^{-2}$:

$$D^B(\beta) = \frac{dB}{d\beta} = (1-B)\left( <E> - 2\frac{<m^2 E>}{<m^2>} + \frac{<m^4 E>}{<m^4>} \right). \qquad (7)$$

4. In addition the specific heat

$$C_V = \beta^2 N_1 \left( <e^2> - <e>^2 \right) \quad \text{with} \quad e = E/N_1 \qquad (8)$$

is considered. The maximum values of these quantities are denoted by $\chi_c(N_0)$, $D_c^m(N_0)$, $D_c^B(N_0)$ and $C_{V,c}(N_0)$. Their FSS behavior determines the critical exponents $\gamma$, $\beta$, $\nu$ and $\alpha$ through the following equations:

$$\chi_c = a_{11} + a_{12} N_0^{\frac{\gamma}{2\nu}}, \quad D_c^m = a_{21} + a_{22} N_0^{\frac{1-\beta}{2\nu}}, \qquad (9a)$$

$$D_c^B = a_{31} + a_{32} N_0^{\frac{1}{2\nu}}, \quad \text{and} \quad C_{V,c} = a_{41} + a_{42} N_0^{\frac{\alpha}{2\nu}}. \qquad (9b)$$

Here we have neglected higher order corrections in $N_0$. The expected exponents, KPZ versus Onsager, are collected in table 2. As $\alpha \leq 0$, the specific heat does not really define a pseudocritical point. In the $N_0 \to \infty$ limit, the pseudocritical points converge towards the critical point:

$$\beta_c^i(N_0) = \beta_c + b_i N_0^{-\frac{1}{2\nu}}. \qquad (10)$$

We employ the multicanonical method [14] to cover for each $N_0$ a $\beta$-range large enough to accomodate all pseudocritical points in one run. An additional advantage is that metastabilities may be avoided through excursions into the disordered phase.

## 2.3 Numerical results

Tables 3 and 4 give an overview of our statistics and simulation results for geometry $G1$ and $G2$. The statistics is given in sweeps, where one sweep is defined by updating in the average $1/2$ of the links and $3/2$ of the spins. Measurements were performed every ten sweeps, except for $N_0 = 4500$ for which they were performed every twenty sweeps. Error bars are calculated with respect to twenty jackknife bins. They are given in parenthesis and apply to the last two given digits.

Figure 2 shows our multicanonical energy distribution for the $N_0 = 4500$ lattice. This should be compared with figure 3, where we show the functions $\chi(\beta)$, $15 D^m(\beta)$ and $60 C_V(\beta)$ on a semi-log scale (the factors are chosen such that all quantities can be depicted together). It is obvious from figure 3 that $\chi_c$ should allow a more accurate FSS analysis than $D_c^m$. The maximum of the specific heat shifts with $N_0$, and stays essentially constant. Figure 4 depicts $D^B(\beta)$. Due to the higher moments in its definition (7) substantially larger relative errors than in figure 3 are encountered.

We now turn to the FSS analysis. For geometry $G1$ we have the larger statistics due to simulating an $N_0 = 4500$ lattice. Nevertheless, we face the problem that three parameter fits to the equations (9), are unstable. Substantially larger lattices would be needed, but their simulation is impossible within our present CPU time limitations. Therefore, our analysis has to rely on two parameter fits. We assume exact KPZ or Onsager exponents, see table 2, and ask the question whether consistent fits are then possible. To decide it we have calculated the goodness-of-fit [15] $Q$. Under the assumption that the discrepancy between the data and the fit function is entirely due to statistical fluctuations of the data, $Q$ is an uniformly distributed random variable in the range $(0:1]$. Our fits rely on lattices with $N_0 \geq 500$, as the $N_0 = 250$ lattice turned out to be too small to exhibit asymptotic (large $N_0$) behavior. This means, we use six data points for geometry $G1$ and five data points for geometry $G2$. The results for $Q$ are collected in table 5. It is clear that Onsager exponents are ruled out, the likelihood that the encountered discrepancy is due to chance is apparently $< 10^{-25}$ for geometry $G1$ and $< 2 \cdot 10^{-4}$ for $G2$. Only the geometry $G2$ fit for $1/(2\nu)$ is consistent with the Onsager exponent, presumably due to the large statistical errors of the corresponding data. On the other hand, the data are well consistent with KPZ exponents. There is one $Q = 0.04$ value, but with six independent observables the likelihood to encounter this value would be about 20%, *i.e.* their is no statistically significant discrepancy. Figure 5 compares KPZ versus Onsager fits for $[\chi_c(N_0) - \chi_1(N_0)]$, where $\chi_1(N_0)$ is the geometry $G1$ KPZ fit for $\chi_c(N_0)$.

Assuming now the KPZ value for $\nu$, equation (10) is used to estimate the critical $\beta$. Smaller lattices are successively eliminated from the fits, until an acceptable $Q$ value is obtained. The thus obtained $\beta_c$ estimates are also included in table 5. For geometry $G1$ the estimates indicate that systematic errors exceed the statistical error significantly. It is not entirely clear whether it would need larger lattices to reconcile these values. Possibly the observables, in particular the derivative of the Binder cumulant, suffer from systematic errors which one could improve by just enlarging the statistics. For $G2$ we do not find such an inconsistency, but there is also a problem with the $D_c^B$ data: the goodness-of-fit $Q$ gets only acceptable when the fit is constrained to the last three lattices. Six data points were used for all other $\beta_c$ fits. Taking weighted averages we get

$$\beta_c = 1.123 \pm 0.002 \ \ \text{for} \ \ G1 \ \ \text{and} \ \ \beta_c = 1.120 \pm 0.005 \ \ \text{for} \ \ G2. \tag{11}$$

Here the error bar is taken from the $\chi_c$ estimates, as other estimates are from the same configurations, it would be inappropriate to use them for error bar reduction.

Data for the specific heat follow in table 6. In addition to the maximum values $C_{V,c}$ we also give the values at the infinite volume critical points: $C_V(\beta_c)$ with $\beta_c$ estimated by equation (11). The Onsager value $\alpha = 0$ would imply a logarithmic behavior, and we use the fit

$$C_{V,c} = a_{41} + a_{42} \ln(N_0)$$

to compare it with the fit (9b) for KPZ exponents. The $Q$ values for these fits are collected in the last two rows of table 6. From the maximum values $C_{V,c}$ we cannot distinguish between KPZ and Onsager. As one may have expected, both fits are consistent, presumably due to a dominant regular part. Somewhat surprising is the bad quality of the same fits when $C_V(\beta_c)$ instead of $C_{V,c}$ is used. Onsager becomes then inconsistent. Although some orders of magnitude better, the quality of the KPZ fits is not too convincing either. Most likely the reason is that the asymptotic $C_V(\beta_c)$ behavior will set it in only for larger lattices. Assuming now $\alpha = -1$, it should be noted that there is now reason for the location of the maximum to approach $\beta_c$. Indeed, changing the multiplicative factor $\beta^2$ in equation (8) to another power would shift the position of the maximum. This is the reason, why we also listed $C_V(\beta_c)$ and do not attempt to estimate $\beta_c$ by means of FSS analysis of the $\beta_c^4(N_0)$.

Another set of two parameter fits is obtained by setting the constants $a_{i1}$ in equations (9) equal to zero. This gives direct estimates of the critical exponents involved, which for sufficiently accurate data from large enough lattices will converge to the exponents of the infinite volume theory. We include the $N_0 = 250$ lattice, if the goodness-of-fit is already satisfactory, otherwise, we apply the fits to our $N_0 \geq 500$ lattices. The results are collected in table 7. They are a bit off from KPZ exponents, although closer than typical results from SDT simulations [8, 9, 10, 11]. The fact that we can perform different two parameter fits, which are both consistent, explains why three parameter fits are still unstable. In the investigated range of lattices the constants $a_{i1}$ are still important, in particular for geometry $G2$.

We have also started to investigate geometries on the torus. Apparently this is hampered by the fact that regular initial configurations give rise to a substantial fraction of flips which result in degenerate triangles of area zero. In the present paper we have excluded these flips, but this is harmless as for the considered geometries on the sphere they are of measure zero. From the torus we have preliminary indications that there may be geometries which belong to yet different universality classes. The basic problem remains, to disentangle gravitational dressing (if any) from fixed geometry DT interactions. Future investigations may provide more insight.

# 3 Interpretation

It is well known that the critical properties of the 2D Ising model correspond to a free fermion theory. It is natural to assume that (fixed geometry) DT moves will give rise to effective four fermion interactions. The anomalous dimension $d_{\overline{\psi}\psi}$, where $\psi$ is is the massless interacting Thirring field, allows then to parametrize the critical exponents [16]:

$$\nu = (2 - d_{\overline{\psi}\psi})^{-1} \text{ and } \eta = d_{\overline{\psi}\psi}/2. \tag{12}$$

The other exponents follow through the scaling laws

$$1 - \alpha/2 = \nu, \ \alpha + 2\beta + \gamma = 2 \text{ and } \gamma = (2-\eta)\nu,$$

which in this model are easily seen to hold. The choice $d_{\overline{\psi}\psi} = 4/3$ does indeed reproduce the KPZ exponents. Our conjecture is that for a fixed geometry DT moves induce the appropriate four fermion interaction. Introducing then "gravitational" effects through SDT fluctuations of the geometry (*i.e.* setting the $l_{ij} = 1$ and thus $A_{l^T}/(l^T)^2 = \sqrt{3}/4$) has no further effects on the critical exponents.

SDT is considered a promising approach towards the quantization of gravity because it is conjectured [17, 18] to sample representative geometries. A major factor of support towards this conjecture has been the agreement with the exact KPZ exponents. As our results suggest that these exponents are not sensitive to fluctuations of the 2D geometry, the agreement between SDT and KPZ becomes trivial. Any sampling over geometries will do, as long as DT updates are used. A fresh look at the issue whether SDT will sample representative geometries becomes legitimate.

We like to raise the concern that there are geometries which do not allow for smooth approximation by triangulation with equilateral triangles, even in the limit $N_0 \to \infty$. We consider as observable the deficit angle $\delta[p]$ associated with a closed path $p$. As is well known, any vector parallel transported around a closed path will in general undergo a rotation, such that the initial vector and the final vector differ by a deficit angle $\delta[p]$. (In 2D $\delta[p]$ is the same for all kinds of vectors transported along the path $p$ and describes directly the curvature of the enclosed region.) A closed path on a simplicial lattice encircles in general $N[p]$ vertices and $\delta[p]$ is the sum of deficit angles associated with these sites, $\delta[p] = \sum_{i \subset p} \delta_i$, see figure 6 for $N[p] = 2$. Now, $\delta_p = n_p \theta$, where $\theta = \pi/3$ is the dihedral angle of an equilateral triangle, and $n_p$ being an integer (including zero). I.e. $\delta_p$ obtains values in the discrete set $\{... -\frac{2}{3}\pi, -\frac{1}{3}\pi, 0, \frac{1}{3}\pi, \frac{2}{3}\pi, ...\}$ only. Clearly, there are many smooth geometries that contain closed paths with associated deficit angles that are not in this set.

Related, there is already no smooth approximation of the 2D sphere by a triangulation with equilateral triangles. The total number of links $N_1$ in the lattice equals $N_1 = 3N_0 - 6$, with $N_0$ being the total number of vertices. On the other hand $N_1 = \frac{1}{2} \sum_i n_i$, where $n_i$ denotes the coordination number of vertex $i$. This gives

$$\sum_i n_i = 6N_0 - 12, \tag{13}$$

and implies restrictions for the deficit angles. For each vertex the deficit angle (*i.e.* the curvature) is simply given as $\delta_i = (6 - n_i)\theta$. Now the positive curvature of the 2-sphere *at every point* cannot be realized with equilateral triangles since $n_i < 6$ is not possible for all vertices due to equation (13). Even if we try $n_i \leq 6$ for each vertex, we see that (almost) all vertices have to have coordination number 6 and a maximum of 12 vertices might show positive curvature. To distribute vertices of positive curvature uniformly on the 2-sphere one has to face as many vertices of negative curvature. As an alternative we could collect all negative curvature in few (singular) vertices.

It seems that one needs ensemble averages over equilateral triangulations to reproduce classical properties of geometries as simple as the 2D sphere. It is then difficult to imagine that fluctuation can be suppressed, for instance how to get a smooth sphere and Onsager exponents in the limit of no gravitational fluctuation? What are small fluctuations in the SDT framework? We think that even in 2D the SDT concept needs to be elaborated in much more details.

## 4  Summary and Conclusions

Two dimensions may be a bad testing ground for quantum gravity concepts. Beautiful exact solutions exist. However, there may be reasons for their existence which obstruct their physical relevance. In essence, two dimensional gravity seems to be solvable, because (a) the classically dominant Einstein term does not fluctuate, and (b) there are (hidden) relations to solvable models in flat space. Therefore, we consider it unlikely that it will provide valuable guidance towards realistic quantum gravity problems. We have illustrated this point, by demonstrating consistency with *no* gravitational dressing, once the proper fixed geometry model is used as starting point to include the effects of geometry fluctuations. It seems that SDT relates to fixed geometry DT spin systems in a similar way as the Regge approach relates to the Onsager limit. Of course, numerical results cannot prove that the fixed geometry DT exponents are really identical with the KPZ exponents. But certainly our results show that fixed geometry DT introduces non-trivial interactions which lead out of the Onsager universality class.

Acknowledgements: On of the authors (BB) would like to thank Herbert Hamber for drawing our attention to Ref. [7].

## References

[1] V.A. Kasakov, JETP Lett. 44, 133 (1986); Phys. Lett. A119 (1986) 140.

[2] D.V. Boulatov and V.A. Kasakov, Phys. Lett. B186 (1987) 379.

[3] V.G. Knizhnik, A.M. Polyakov, and A.B. Zamalodchikov, Mod. Phys. Lett. A3 (1988) 819.

[4] A.A. Migdal, Nucl. Phys. B (Proc. Suppl.) 9 (1989) 625.

[5] M. Gross and H. Hamber, Nucl. Phys. B364 (1991) 703.

[6] C. Holm and W. Janke, Phys. Lett. B335 (1994) 143.

[7] M. Vekić, S. Liu, and H. Hamber, Phys. Lett. B329 (1994) 444, and hep-lat 9407166 (1994). As the flat space weight factors are set identical to one in this work, we think that a more appropriate interpretation would be that of a SDT simulation with a cuttoff.

[8] J. Jurkiewicz, A. Krzywicki, B. Petersson and B. Söderberg, Phys. Lett. B213 (1988) 511.

[9] S.M. Cateral, J.B. Kogut and R.L. Renken, Phys. Rev. D45 (1992) 2957.

[10] C. Baillie and D. Johnston, Mod. Phys. Lett. A7 (1992) 1519.

[11] R. Ben-Av, J. Kinar and S. Solomon, Int. J. Mod. Phys. C3 (1992) 279.

[12] J.W. Alexander, Ann. Math. 31 (1930) 292.

[13] K. Binder, Z. Phys. B43 (1981) 119.

[14] B. Berg and T. Neuhaus, Phys. Lett. B267 (1991) 249.

[15] W.H. Press, B.P. Flannery, S.A. Teukolsky, and W.T. Vetterling, *Numerical Recipes*, (Cambridge University Press, 1988).

[16] B. Berg, Annal. Phys. 110 (1978) 329; 122 (1979) 463.

[17] V.A. Kazakov, I.K. Kostov and A.A. Migdal, Phys. Lett. 157B (1985) 295.

[18] J. Ambjorn, Les Houches Lectures, (1994).

# Tables

| $n$ | 3 | 4 | 5 | 6 | 7 | 8 | 9 | 10 | 11 | 12 | 13 | 14 | 15 | 16 | 17 | 18 | | 26 |
|---|---|---|---|---|---|---|---|---|---|---|---|---|---|---|---|---|---|---|
| $G1$ | 1 | 69 | 49 | 57 | 43 | 13 | 5 | 1 | 6 | 1 | 1 | 1 | 1 | 1 | 0 | 0 | | 1 |
| $G2$ | 0 | 62 | 81 | 34 | 26 | 17 | 14 | 5 | 5 | 2 | 2 | 1 | 0 | 0 | 0 | 1 | | 0 |

Table 1: *Distribution of coordination numbers n for geometries G1 and G2 in case of $N_0 = 250$ sites.*

| Exponent | KPZ | Onsager |
|---|---|---|
| $\gamma/(2\nu)$ | $0.66\overline{6}$ | 0.875 |
| $(1-\beta)/(2\nu)$ | $0.166\overline{6}$ | 0.4375 |
| $1/(2\nu)$ | $0.33\overline{3}$ | 0.5 |
| $\alpha/(2\nu)$ | $-0.33\overline{3}$ | 0 |

Table 2: *KPZ versus Onsager critical exponents.*

| $N_0$ | Sweeps | $\chi_c$ | $\beta_c^1$ | $D_c^m$ | $\beta_c^2$ | $D_c^B$ | $\beta_c^3$ |
|---|---|---|---|---|---|---|---|
| 250 | $4 \cdot 10^6$ | 28.60 (12) | 1.1819 (13) | 1.6290 (50) | 1.2096 (14) | 1.515 (14) | 1.0875 (32) |
| 500 | $4 \cdot 10^6$ | 46.53 (25) | 1.1720 (09) | 2.0253 (86) | 1.1968 (12) | 1.916 (23) | 1.1027 (31) |
| 1000 | $4 \cdot 10^6$ | 73.03 (51) | 1.1587 (11) | 2.400 (16) | 1.1806 (15) | 2.450 (33) | 1.1098 (19) |
| 1500 | $4 \cdot 10^6$ | 95.75 (79) | 1.1560 (11) | 2.620 (21) | 1.1773 (16) | 2.741 (49) | 1.1031 (53) |
| 2000 | $6.7 \cdot 10^6$ | 115.3 (1.0) | 1.1546 (12) | 2.832 (25) | 1.1739 (13) | 2.918 (48) | 1.1092 (36) |
| 3000 | $10^7$ | 147.3 (1.3) | 1.1481 (11) | 3.039 (16) | 1.1665 (16) | 3.324 (45) | 1.1028 (26) |
| 4500 | $7 \cdot 10^6$ | 194.6 (2.1) | 1.1463 (11) | 3.349 (36) | 1.1627 (19) | 3.827 (90) | 1.1045 (35) |

Table 3: *Statistics and numerical results for geometry G1.*

| $N_0$ | Sweeps | $\chi_c$ | $\beta_c^1$ | $D_c^m$ | $\beta_c^2$ | $D_c^B$ | $\beta_c^3$ |
|---|---|---|---|---|---|---|---|
| 250 | $4 \cdot 10^6$ | 28.65 (12) | 1.2138 (22) | 1.2907 (44) | 1.2094 (31) | 1.265 (15) | 1.0479 (41) |
| 500 | $4 \cdot 10^6$ | 44.24 (33) | 1.1934 (25) | 1.4491 (89) | 1.1899 (41) | 1.533 (38) | 1.004 (15) |
| 1000 | $4 \cdot 10^6$ | 67.98 (85) | 1.1769 (26) | 1.601 (13) | 1.1753 (46) | 1.736 (51) | 1.0301 (75) |
| 1500 | $4 \cdot 10^6$ | 85.4 (1.7) | 1.1722 (53) | 1.647 (19) | 1.175 (12) | 1.923 (52) | 1.0051 (71) |
| 2000 | $6.7 \cdot 10^6$ | 102.5 (1.6) | 1.1719 (41) | 1.690 (17) | 1.1770 (89) | 1.991 (35) | 1.032 (14) |
| 3000 | $10^7$ | 133.2 (1.8) | 1.1670 (74) | 1.758 (15) | 1.1493 (50) | 2.104 (28) | 1.022 (12) |

Table 4: *Statistics and numerical results for geometry G2.*

| Observable | $G1$ KPZ | $G1$ Onsager | $G1$ $\beta_c$ | $G2$ KPZ | $G2$ Onsager | $G2$ $\beta_c$ |
|---|---|---|---|---|---|---|
| $\chi_c(N_0)$ | 0.32 | $10^{-25}$ | 1.1226 (17) | 0.66 | $2 \cdot 10^{-4}$ | 1.1227 (43) |
| $D_c^m(N_0)$ | 0.52 | $10^{-4}$ | 1.1319 (26) | 0.04 | $7 \cdot 10^{-4}$ | 1.1135 (73) |
| $D_c^B(N_0)$ | 0.20 | $10^{-3}$ | 1.0961 (53) | 0.43 | 0.20 | 1.103 (63) |

Table 5: *Goodness-of-fit Q, assuming KPZ versus Onsager critical exponents. In addition $\beta_c$ estimates are given which assume the KPZ exponent $\nu = 1.5$*

| $N_0$ | $G1$ $C_{V,c}$ | $G1$ $C_V(\beta_c)$ | $G1$ $\beta_c^4$ | $G2$ $C_{V,c}$ | $G2$ $C_V(\beta_c)$ | $G2$ $\beta_c^4$ |
|---|---|---|---|---|---|---|
| 250 | 0.29239 (61) | 0.22742 (65) | 1.2670 (20) | 0.25617 (59) | 0.22038 (66) | 1.2779 (49) |
| 500 | 0.30279 (82) | 0.23449 (49) | 1.2413 (22) | 0.25427 (61) | 0.22561 (79) | 1.2569 (65) |
| 1000 | 0.30698 (85) | 0.24291 (99) | 1.2250 (26) | 0.2521 (11) | 0.22802 (81) | 1.2618 (98) |
| 1500 | 0.30585 (84) | 0.2458 (11) | 1.2148 (32) | 0.25337 (86) | 0.2263 (11) | 1.2563 (87) |
| 2000 | 0.3084 (11) | 0.24650 (88) | 1.2108 (41) | 0.25212 (56) | 0.22737 (61) | 1.2325 (99) |
| 3000 | 0.3083 (11) | 0.2499 (10) | 1.2043 (30) | 0.25219 (68) | 0.22875 (48) | 1.2239 (30) |
| 4500 | 0.3089 (11) | 0.25017 (61) | 1.1937 (51) | | | |
| KPZ | 0.31 | 0.09 | | 0.67 | 0.014 | |
| Onsager | 0.19 | $2 \cdot 10^{-3}$ | | 0.47 | $10^{-3}$ | |

Table 6: *Numerical results for the specific heat (geometries G1 and G2) The last two rows give the goodness-of-fits Q for the KPZ and the Onsager scenarios.*

| Exponent | $G1$ | $G2$ |
|---|---|---|
| $\gamma/(2\nu)$ | 6: 0.6489 (43) [0.53] | 6: 0.6185 (42) [0.48] |
| $(1 - \beta)/(2\nu)$ | 6: 0.2211 (36) [0.45] | 5: 0.1082 (54) [0.07] |
| $1/(2\nu)$ | 7: 0.3185 (52) [0.12] | 5: 0.176 (15) [0.66] |

Table 7: *Two parameter fits for the exponents. The number of data used is indicated first, estimates and their error bars follow, finally the number [Q] gives the goodness-of-fit.*

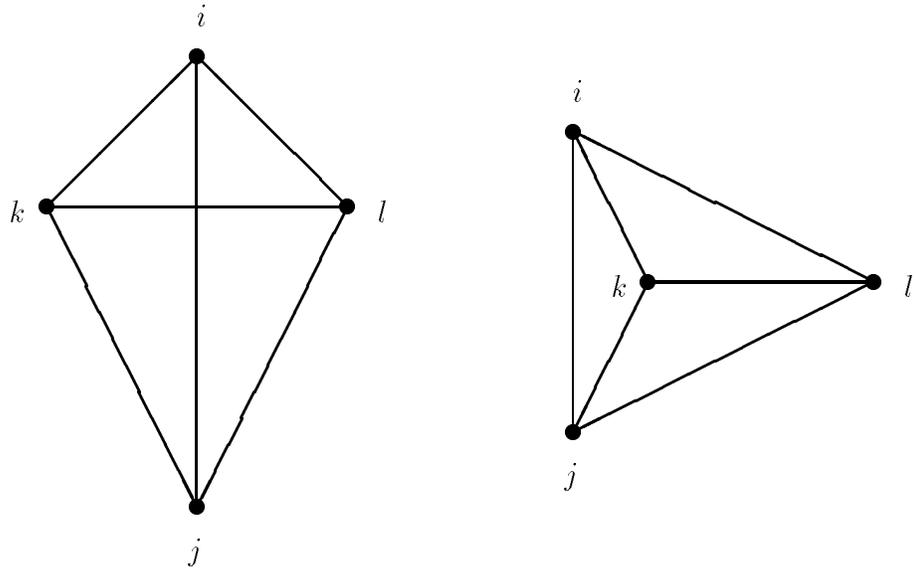

Figure 1: The left picture shows the flip of a link $kl \rightarrow ij$, preserving the geometry. On the right a flip is depicted which is not performed since the geometry changes, the area would increase.

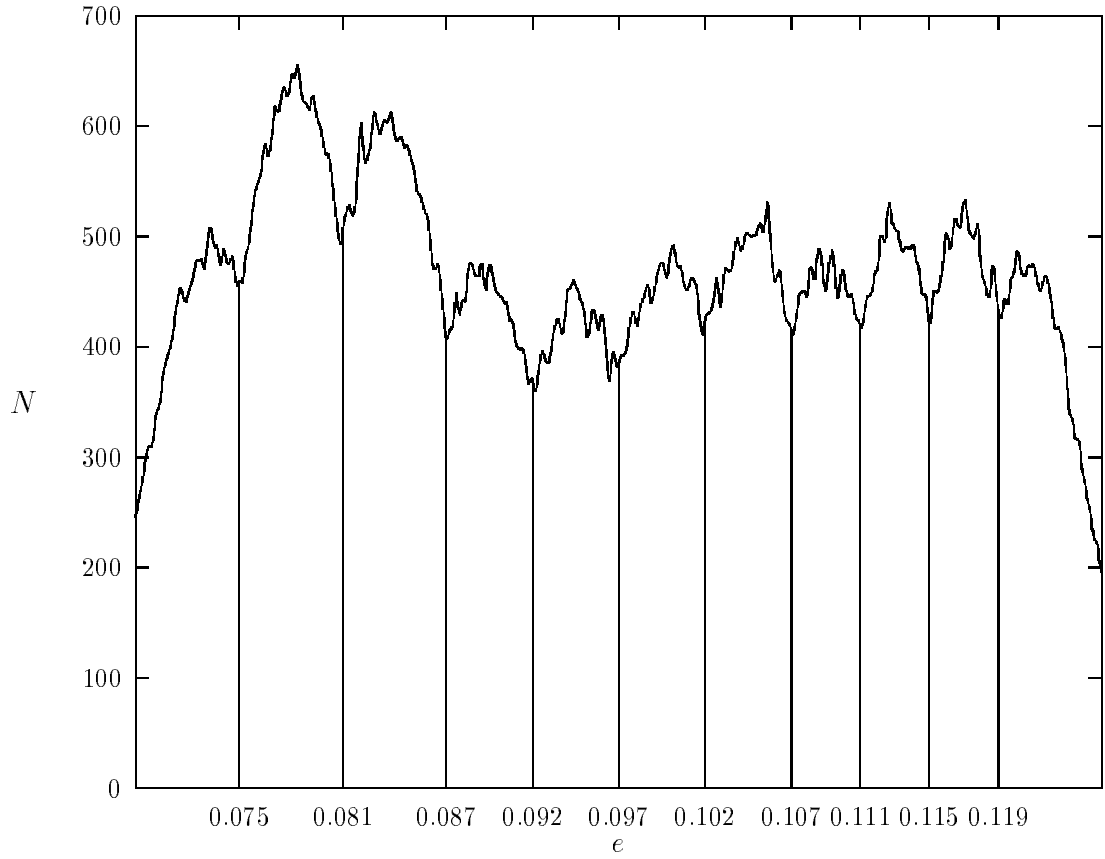

Figure 2: Multicanonical energy distribution for the $G1$, $N_0 = 4500$ lattice. In the indicated interval $\beta$ decreases from 1.25 in steps of 0.025 down to 1.

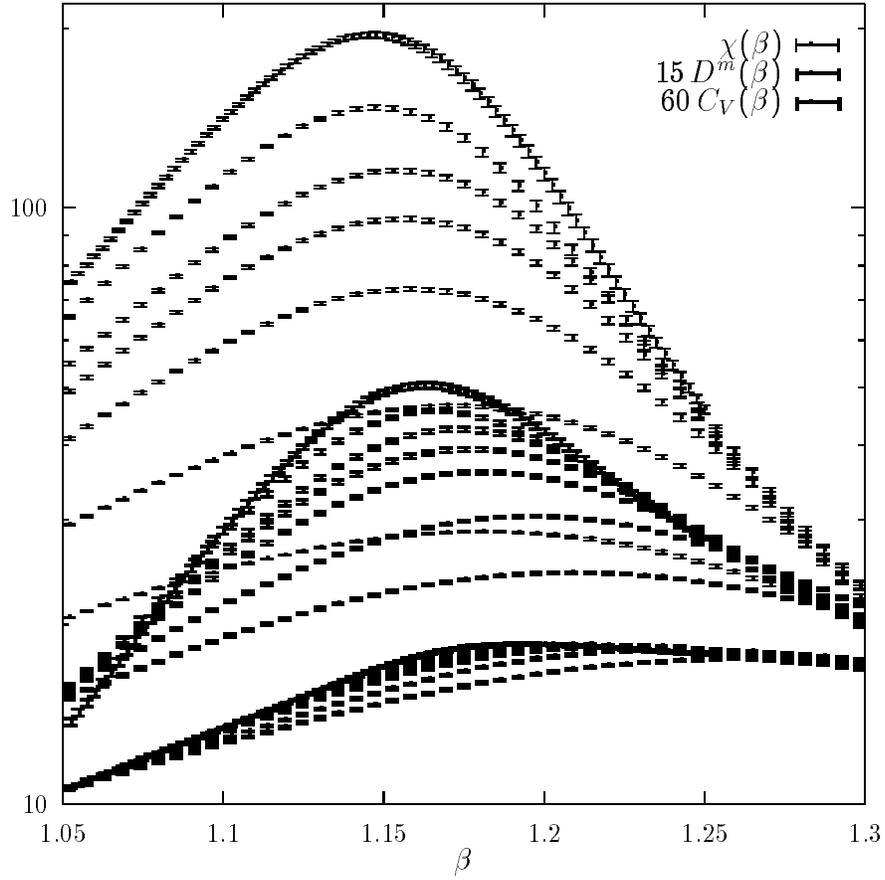

Figure 3: Geometry $G1$: Susceptibility $\chi(\beta)$, derivative of the (absolute) magnetization $D^m(\beta)$ and specific heat $C_V(\beta)$. From up to down the curves correspond (in each case) to $N_0 = 4500, 3000, 2000, 1500, 1000, 500$ and $250$.

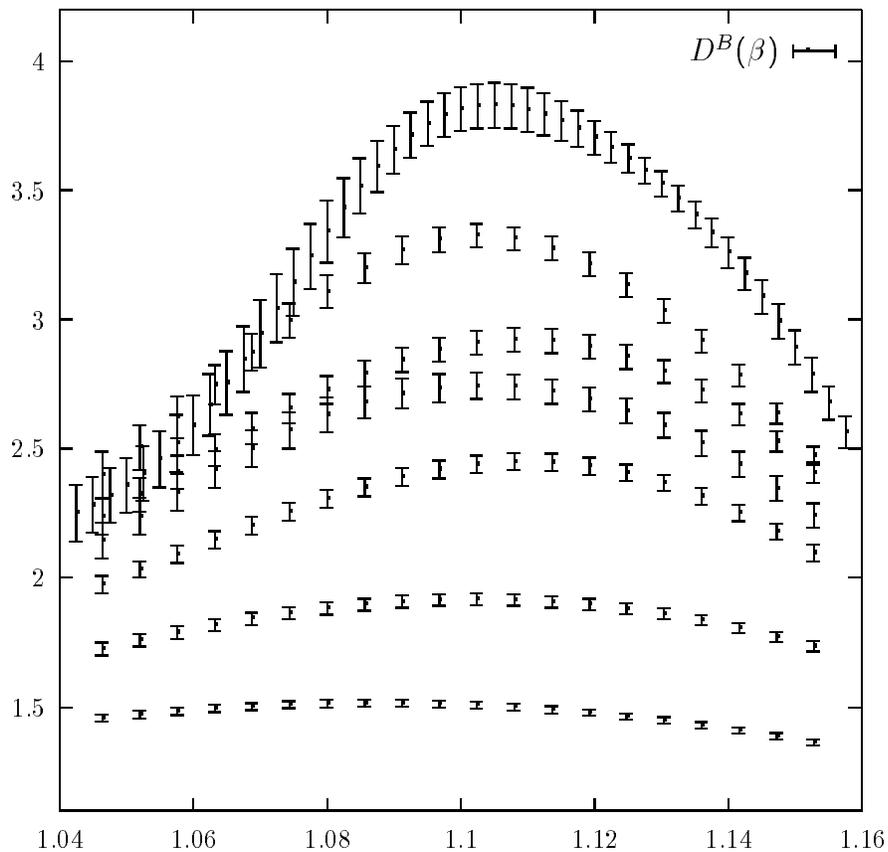

Figure 4: Geometry $G1$: Derivative of the Binder cumulant $D^B(\beta)$. From up to down the data correspond to $N_0 = 4500,\ 3000,\ 2000,\ 1500,\ 1000,\ 500$ and $250$.

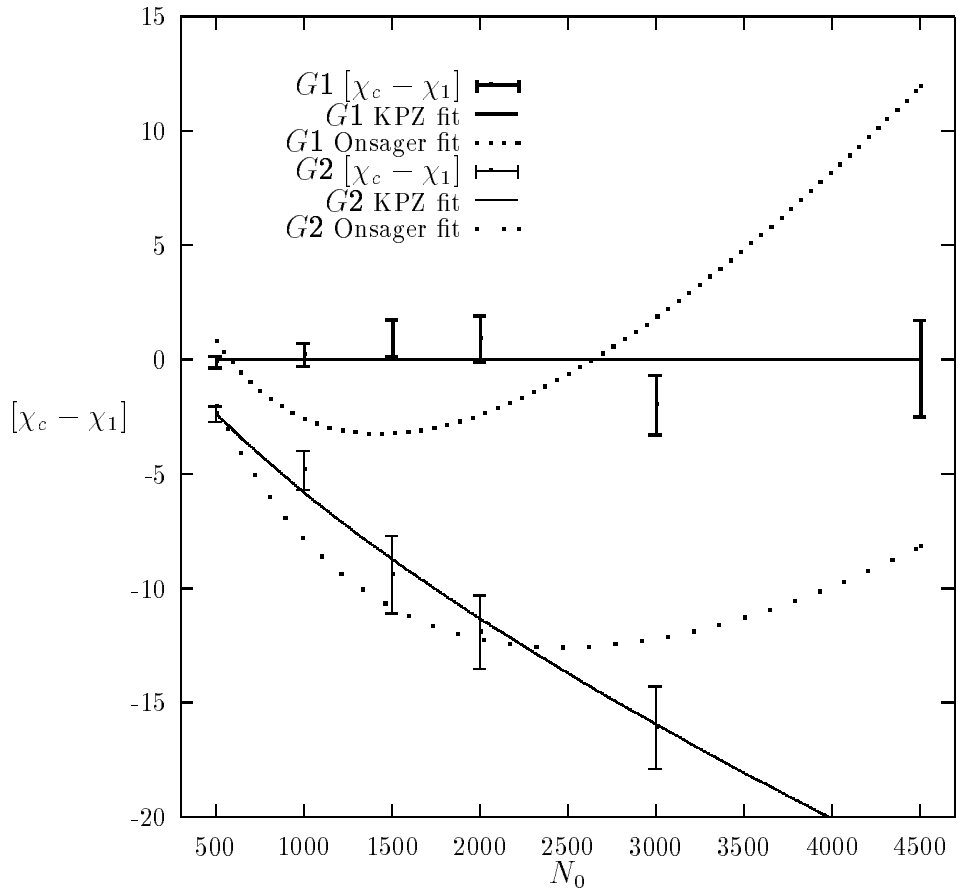

Figure 5: KPZ versus Onsager fits for geometries $G1$ and $G2$. Depicted is $[\chi_c(N_0) - \chi_1(N_0)]$, where $\chi_c$ are the measured maxima of the susceptibility, and $\chi_1$ is the geometry $G1$ KPZ fit for this quantity.

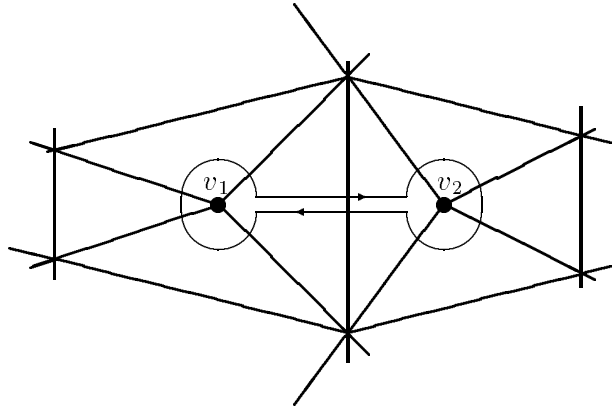

Figure 6: A path encircles two vertices. Any vector transported around this path will be rotated by an angle $\delta_p$. The trip around vertex $v_1$ leads to a rotation $\delta_{v_1}$, followed by an additional rotation caused by the deficit angle $\delta_{v_2}$ of vertex $v_2$. The total deficit angle $\delta_p$ is thus the sum $\delta_{v_1} + \delta_{v_2}$. Stretching the path to encircle more vertices generalizes the result for an arbitrary number of vertices.